\documentclass[11pt,twoside]{article}

\usepackage{asp2006}
\usepackage{epsf}
\usepackage{graphicx}
\usepackage{lscape}
\def\be{\begin{equation}}
\def\ee{\end{equation}}

\def\ergs{{\rm\,erg\,s^{-1}}}
\def\msun{M_{\odot}}

\def\ergs{\rm \,erg\,s^{-1}}
\catcode`\@=11 
\def\@versim#1#2{\vcenter{\offinterlineskip
        \ialign{$\m@th#1\hfil##\hfil$\crcr#2\crcr\sim\crcr } }}
\def\lsim{\mathrel{\mathpalette\@versim<}}
\def\gsim{\mathrel{\mathpalette\@versim>}}
\def\mpy{M_\odot \ {\rm yr^{-1}}}

\begin{document}

\title{Accretion and ejection in Sgr A*}
\author{Feng Yuan}   
\affil{Shanghai Astronomical Observatory, Chinese Academy of
Sciences, 80 Nandan Road, Shanghai 200030, China; fyuan@shao.ac.cn}

\begin{abstract} 

We review our current understanding to the accretion and ejection
processes in Sgr A*. Roughly speaking, they correspond to the
quiescent and flare states of the source respectively. The
high-resolution {\it Chandra} observations to the gas at the Bondi
radius combined with the Bondi accretion theory, the spectral energy
distribution from radio to X-ray, and the radio polarization provide
us strict constraints and abundant information to the theory of
accretion. We review these observational results and describe how
the advection-dominated accretion flow model explains these
observations. Recently more attentions have been paid to flares in
Sgr A*. Many simultaneous multi-wavelength campaigns have been
conducted, aiming at uncovering the nature of flares. The main
observational properties of flares are briefly reviewed. Especially,
the time lag between the peaks of flare at two radio frequencies
strongly indicates that the flare is associated with ejection of
radio-emitting blobs from the underlying accretion flow. Such kind
of episodic jets is distinctive from the continuous jets and are
quite common in black hole systems. We introduce the
magnetohydrodynamical model for the formation of episodic jets
recently proposed based on the analogy with the theory of coronal
mass ejection in the Sun. We point out that the various
observational appearances of flares should be explained in the
framework of this model, since ejection and flare originate from the
same physical process.

\end{abstract}

\section{Introduction}

Accretion onto compact objects is one of the most fundamental
processes in the universe. It operates in various scales, ranging
from the centers of almost each galaxies including our own, black
hole and neutron star X-ray binaries, Gamma-ray bursts, to
protostars. Among them the suppermassive black hole in our Galaxy,
Sgr A*, is unique because of its proximity (Sch\"odel et al. 2002;
Ghez et al. 2003). Sgr A* is one of the most strongest evidence for
the existence of supermassive black holes. Because of this reason we
have tremendously abundant data in various aspects, as we will
describe below. This supplies us the most strict constraint to
theoretical models.

In this review, we will focus on two aspects of Sgr A*. One is its
quiescent state. The corresponding theory is black hole accretion.
We will introduce the development of the advection-dominated
accretion flow (ADAF) and illustrate how the main observations of
the quiescent state of Sgr A* are naturally explained in this model.
In fact, since its re-discovery in 1994, the model has been
intensively studied via analytical work, numerical simulation, and
comparison with observations, so the ADAF model has been well
established and become the standard model of low luminosity sources.

Another aspect we want to focus is flares of Sgr A*. This is a ``hot
spot'' in recent years in Sgr A* community. Many simultaneous
multi-waveband campaigns aiming at studying flares have been
conducted successfully and valuable information have been obtained
from them. As we will introduce below, these observations combined
with some convincing theoretical works clearly indicate that the
flares are physically associated with separate blob-like mass
ejection from the accretion flow. We want to emphasize that this
kind of mass ejection is different from the continuous jets we
usually talk about. Rather, they are usually called ``episodic
jets''. Although the theoretical study of episodic jets is still in
its infancy compared to the continuous jets, they are actually very
common in black hole systems. We will introduce an MHD model for the
formation of episodic jets which is recently proposed by analogy
with the coronal mass ejection in the Sun (Yuan et al. 2009). We
argue that flares in Sgr A* should be explained in the framework of
this model.

\section{Quiescent state of Sgr A*: the baseline model}

\subsection{Why unique? Observational constraints}

The reason why Sgr A* is unique is because of its proximity. We
therefore have abundant data which provides strict observational
constraints on accretion models. Below we state the main
constraints.

{\em Outer boundary conditions}. The accretion flow starts at the
Bondi radius where the gravitational potential energy of the gas
equals its thermal energy. For uniformly distributed matter with an
ambient density $\rho_0$ and sound speed $c_s$, the Bondi radius of
a black hole with mass $M$ is $R_{\rm Bondi}\approx GM/c_s^2$ and
the mass accretion rate is $\dot{M}_{\rm Bondi}\approx 4\pi R_{\rm
Bondi}^2\rho_0c_s$. The high spacial resolution of {\em Chandra} can
well resolve the Bondi radius and these observations infer gas
density and temperature as $\approx 100~ {\rm cm}^{-3}$ and $\approx
2 $ keV on $1^{\arcsec}$ scales (Baganoff et al. 2003). The
corresponding Bondi radius $R_{\rm Bondi} \approx 0.04 {\rm
pc}\approx 1^{\arcsec}\approx 10^5R_s$ and $\dot{M}_{\rm Bondi}
\approx 10^{-5} \mpy$. The 3D numerical simulation for the accretion
of stellar winds onto Sgr A* by Cuadra et al. (2006) obtains
$\dot{M}\approx 3\times 10^{-6}\mpy$, in good agreement with the
estimation of the simple Bondi theory. This simulation also tells us
that the angular momentum of the accretion flow at Bondi radius is
not small, with the circularization radius being $\sim 10^4R_s$. The
Bondi radius, Bondi accretion rate, and the density, temperature,
and the specific angular momentum of the gas at the Bondi radius
constitute the outer boundary conditions that any accretion models
must satisfy.

{\em Spectral energy distribution}. The spectral energy distribution
of Sgr A* is shown in Fig. 1. The figure is directly taken from
Yuan, Quataert \& Narayan (2003), so some later observational data
such as Genzel et al. (2003) and Sch\"odel et al. (2007) are
missing, but they do not affect our discussions here. The radio
spectrum consists of two components. Below $\sim 86$ GHz, the
spectrum is described by a power-law form, while above this
frequency we have a so-called ``sub-millimeter bump''. This implies
that the radio emission comes from two different components. The
infrared (IR) observations always show some variations which seems
to indicate that there is no real quiescent state in IR band (Genzel
et al. 2003; Ghez et al. 2004). The X-ray emission comes in two
states, namely quiescent and flare ones, with different spectral
index, as shown in Figure 1. A large fraction of the X-ray flux in
the quiescent state is resolved, coming from an extended region. The
bolometric luminosity of Sgr A* is only $L\approx 10^{36}\ergs
\approx 3\times 10^{-9}L_{\rm Edd}$. Combined with the Bondi
accretion rate, this luminosity implies an extremely low radiative
efficiency, $\eta\sim 10^{-6}$.

{\em Polarization}. A high level of linear polarization
($2\%\sim10\%$) at frequencies higher than $\sim 150$ GHz was
detected (e.g., Aitken et al. 2001; Bower et al. 2003, 2005;
Macquart et al. 2006; Marrone et al. 2007), which sets the mean
rotation measure to be $-5.6\pm0.7\times 10^5 {\rm rad~m^{-2}}$
(Marrone et al. 2007). This limits the accretion rate close to the
black hole to less than $2\times 10^{-7} \msun {\rm yr}^{-1}$ or
$2\times 10^{-9}\msun {\rm yr}^{-1}$, depending on the configuration
of the magnetic field in the accretion flow\footnote{Ballantyne,
\"Ozel, \& Psaltis (2007) argue that since the electrons are
relativistic over a large radii, polarized radiation from accretion
flow will undergo generalized Faraday rotation, which will not
depolarize the radiation even if the rotation measure is very high.
Thus the polarization observations will not be able to constrain the
accretion rate. However, the electrons are actually non-relativistic
at radii larger than $\sim 100 r_s$ where most of the rotation
measure is produced. On the observational side, the observed
polarization angle varies as $\lambda^2$ (e.g., Macquart et al.
2006), which is what expected for Faraday rotation.}. Such a mass
accretion rate is much lower than the Bondi rate.

\subsection{The standard thin disk is ruled out}

The Bondi theory provides a good estimation to the accretion rate at
the outer boundary. If the gas were accreted at this rate onto the
black hole via the standard thin disk, the expected luminosity would
be $L\approx 0.1\dot{M}_{\rm Bondi} c^2\approx 10^{41}\ergs$, five
orders of magnitude higher than the observed luminosity. This is the
strongest argument against a standard thin disk in Sgr A*. The
second argument against this model is that the spectrum shown in
Fig. 1 does not look like the multi-temperature blackbody spectrum
predicted by a standard thin disk.

\subsection{The advection-dominated accretion flow model}

The equations of accretion onto a black hole allow two series of
solutions, namely cool and hot ones. The standard thin disk model is
the representative of the cool solution (another cool accretion
solution corresponds to higher accretion rates and is ``slim''
disk). The temperature of the accretion flow in this solution is
relatively low, $\sim 10^6-10^7$K. It is optically thick,
geometrically thin (because of the low temperature), and radiatively
efficient. This solution provides a good description to the big blue
bump in the optical/UV band of the luminous AGNs (but see Koratkar
\& Blaes 1999) thus is believed to work in luminous AGNs.

The advection-dominated accretion flow belongs to the hot series
(Narayan \& Yi 1994, 1995; Abramowicz et al. 1995; see Narayan,
Mahadevan \& Quataert 1998 and Narayan \& McClintock 2008 for
reviews). In an ADAF, the temperature of ions is virial while the
electron temperature is lower but still very high, $T_e\sim
10^9-10^{11}$K. This solution is optically thin and geometrically
thick. Most importantly, its radiative efficiency is typically much
lower than that of the standard thin disk, $\eta_{\rm ADAF} \approx
0.1 \dot{M}/\dot{M}_{\rm crit}$. Here $\dot{M}_{\rm crit} \approx
\alpha^2\dot{M}_{\rm Edd}$ is the critical accretion rate of ADAF
beyond which the ADAF solution fails and is replaced by another hot
solution (``luminous hot accretion flow'').

The crucial point to modeling Sgr A* is that the accretion flow must
be radiatively very inefficient. This is exactly the characteristic
feature of an ADAF, as we emphasize above. In a standard thin disk,
the viscously dissipated energy is radiated away locally, which
results in a high efficiency. But in an ADAF, the radial velocity is
much larger and the temperature is much higher than in a standard
thin disk. Consequently, the density of the accretion flow is much
lower. Therefore, the radiative timescale is much longer than the
accretion timescale, thus most of the viscously dissipated energy is
stored in the accretion flow as its thermal energy rather than being
radiated away. This is the main reason for the low radiative
efficiency of an ADAF. For the details of the dynamics of ADAFs, we
refer the reader to the review of Narayan, Mahadevan \& Quataert
(1998) and Narayan \& McClintock (2008).

\begin{figure}[!ht]
\includegraphics[scale=0.5,angle=0,trim=-80 20 -10 -5]{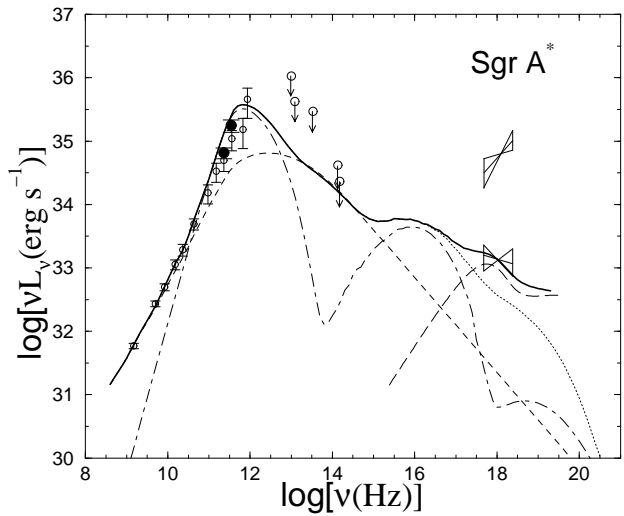}
\caption{ADAF model for the quiescent state emission from Sgr A*.
The dot-dashed line denotes the synchrotron and SSC emission by
thermal electrons; the dashed line for the synchrotron emission by
non-thermal electrons. The dotted line shows the total synchrotron
and SSC emissions while the solid line also includes the
bremsstrahlung emission from the outer parts of the ADAF
(long-dashed line). From Yuan, Quataert \& Narayan (2003).}
\end{figure}

Narayan, Yi \& Mahadevan (1995) first apply the ADAF model to Sgr
A*. They successfully explain the most important feature of the
source, i.e., its low radiative efficiency. The spectrum has also
been roughly explained. However, in the ``old'' ADAF model adopted
in that paper, the mass accretion rate is assumed to be constant
with radius. As a result, the rotation measure predicted in this
model is much larger than that required from the polarization
observation. This is a serious problem.

In the theoretical side, significant progresses on ADAFs have been
made in the past decades. First, global, time-dependent, numerical
simulations reveal that only a very small fraction of the mass
available at large radii actually accretes onto the black hole and
most of it circulates in convective motions or is lost to a
magnetically driven outflow (Stone, Pringle \& Begelman 1999; Hawley
\& Balbus 2002). Second, in the old ADAF model, the turbulent
dissipation is assumed to heat only ions. However, it was later
realized that processes like magnetic reconnection is likely to heat
electrons directly (Quataert \& Gruzinov 1999; Sharma et al. 2007).

Yuan, Quatatert \& Narayan (2003) present a updated ADAF model to
Sgr A*, taking into account the above-mentioned theoretical
developments, i.e., outflow and direct electron heating by turbulent
dissipation. Fig. 1 shows the spectral modeling result.
Specifically, the sub-millimeter bump comes from the synchrotron
emission of thermal electrons in the innermost region of the ADAF.
Due to the convection, only about 1\% of the gas available at the
Bondi radius enters into the black hole horizon\footnote{So the low
efficiency of Sgr A* ($\eta\sim 10^{-6}$) is partly because of
convection ($10^{-2}$) and partly because of energy advection
($10^{-4}$).}, so the density in the region close to the black hole
is much lower than in Narayan et al. (1995). In this case, the
rotation measure is much smaller and a high linear polarization is
expected. The low-frequency radio and the IR emissions are assumed
to be from some nonthermal electrons in the ADAF (since the plasma
in ADAF is collisionless).

After the publication of Yuan et al. (2003), some new observations
appeared. A notable one is the size measurement of Sgr A* at radio
wavebands (Bower et al. 2004; Shen et al. 2005). These results
supply independent test to theoretical models. Yuan, Shen \& Huang
(2006; see also Huang et al. 2007) calculated the predicted size of
Sgr A* of Yuan et al. (2003) model and found good agreement with the
observational results.

\subsection{Comments on the two additional models for Sgr A*}

In addition to the ADAF model, two competitive models are the jet
model (Falcke, Mannheim \& Biermann 1993; Falcke \& Markoff 2000;
Markoff et al. 2001; Yuan, Markoff \& Falcke 2002; Markoff, Bower \&
Falcke 2007), and the spherical accretion model (Melia 1992; Melia,
Liu \& Coker 2000, 2001). Both models can explain the main
observations well. In the jet model, most of the radiation in Sgr A*
comes from synchrotron and synchrotron-self-Compton radiation of
relativistic electrons accelerated in the jet. From a theoretical
point of view, however, compared to accretion flows, there are much
more uncertainties in a jet model in general. The mechanisms of
launching and collimating jets have not been fully understood, thus
large uncertainties exist about the structure of the jet and
magnetic field in it. Regarding the electron acceleration, we are
not sure whether they are accelerated by shock or magnetic
reconnection, and many questions still have not been answered for
both mechanisms. At last, if the radiation of Sgr A* is dominated by
a jet, we have to explain why the radiation of the underlying
accretion flow can be neglected, since the density and radiative
efficiency in the accretion flow is usually higher than in a jet.
Most importantly, so far we have not detected any jet in Sgr A* even
though the proximity of the source and the high sensitivity of the
radio telescope. This is the most serious challenge to the jet
model.

The spherical accretion model has many similar properties with an
ADAF. In both models, the accretion flow is hot, with the
temperature being virial. This is because both models are
advection-dominated. This explains naturally why the radiative
efficiency is so low, as required by observations. In both models,
the main radiative process is the synchrotron emission by thermal
electrons in the accretion flow, which is responsible for the
sub-millimeter bump.

Two key differences exist between the spherical accretion model and
ADAF. The first one is that in the former the specific angular
momentum is assumed to be very low that the flow circularizes at a
radius as small as 5-10 $R_s$. Thus the geometry of the accretion
flow is an outer spherical accretion plus an inner small Keplerian
disk. However, detailed numerical simulations to the accretion
process in the Galactic center region indicates that the specific
angular momentum of the gas is actually not small. The
circularization radius is as large as $10^4 R_s$ (Cuadra et al.
2006). Another difference is that the spherical accretion model
assume a one-temperature plasma while in an ADAF the ions are much
hotter than electrons. Whether the plasma is two-temperature or not
depends on the microphysics of the coupling between ions and
electrons and is theoretically a difficult problem. Strong coupling
mechanism has not been found so far. In this case, adiabatic
compression should results in much hotter ions than electrons due to
the difference of their adiabatic index. On the observational side,
in the cases of supernova remnant and solar wind, the plasma is
found to be two-temperature. At last, recent works seems to show
that it is hard to fit the spectrum of Sgr A* if the accretion flow
is one-temperature (Yuan et al. 2006; Mo\'scibrodzka et al. 2009).

\section{The flare activity and ejection of blobs from Sgr A*}

\subsection{Main properties of flares}

Flares are detected in multi-wavebands, from radio, sub-millimeter,
IR, to X-ray. At both IR and X-ray wavebands, the source is highly
variable. The amplitude of the variability at IR is $\sim$ 1-5
(Genzel et al. 2003; Ghez et al. 2004). Fitting the spectrum with a
power-law, $F_{\nu}\propto \nu^{\alpha}$, Hornstein et al. (2007)
find that the spectral index $\alpha$ of IR flares is
$\alpha=-0.6\pm0.2$, independent of the flux of the flares (but see
Gillessen et al. 2006 for a different result). At X-ray band, the
amplitude of the flare is typically higher, with amplitude as high
as $\sim 45$ or even higher. The spectrum is quite diverse, with
$\alpha=0, 0.1, -0.3, -0.5, -1.5$ (Baganoff et al. 2001; Goldwurm et
al. 2003; Porquet et al. 2003; B\'elanger et al. 2005). The duration
of the IR and X-ray flares are similar, typically 1-3 hours. The
flare event usually occurs a few times per day. It is found that
flares are the more seldom, the more luminous they are (Trippe et
al. 2007; Yusef-Zadeh et al. 2009). In contrast to this, changes in
the radio flux are limited to variations of $<50\%$ within hours to
days (e.g., Trippe et al. 2007; Yusef-Zadeh et al. 2008).

The flare emission is detected to be polarized in radio,
sub-millimeter, and IR bands. In the radio band, compared to the
quiescent state, the polarization degree is higher and the angle
also changes (Yusef-Zadeh et al. 2008). The polarization degree at
sub-millimeter band increases from $9\%$ to $17\%$ as the flare
passes through its peak (Marrone et al. 2008). The IR flare is
strongly polarized, with degree of $20\%$ or even up to $40\%$
(Trippe et al. 2007; Eckart et al. 2008a). This is a strong evidence
for a synchrotron origin of the flare emission.

A quasi-periodicity of $\sim 20$ minutes is claimed in the IR flares
(Genzel et al. 2003; Eckart et al. 2006b; Meyer et al. 2006a, 2006b;
Trippe et al. 2007; Zamaninasab et al. 2010), but this is a subject
of substantial controversy (Meyer et al. 2008; Do et al. 2009).

\subsection{Evidence for ejection of blobs}

\begin{figure}[!ht]
\includegraphics[scale=0.45,angle=0,trim=-60 -20 60 -10]{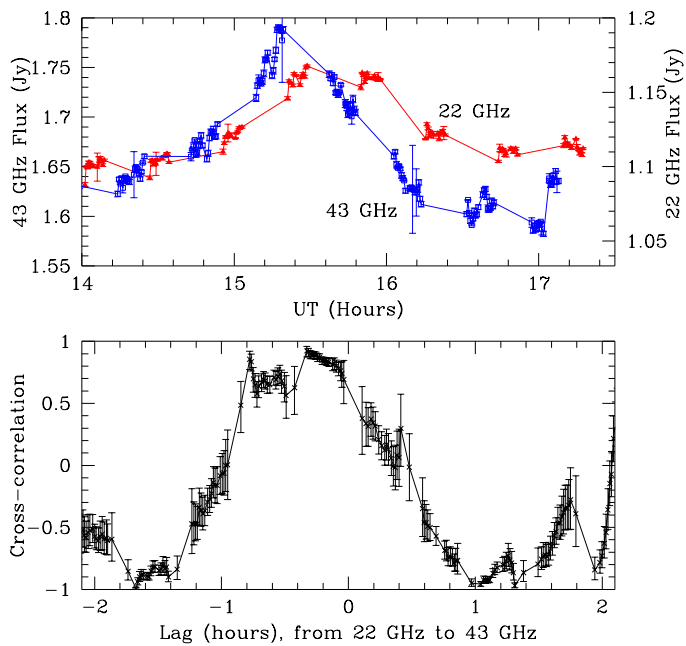}
\caption{{\em Top:} Lightcurve of Sgr A* flare at 43 and 22 GHz.
{\em Bottom:} The cross-correlation amplitude is shown as a function
of lag time. From Yusef-Zadeh et al. (2006b).}
\end{figure}

Many simultaneous multi-waveband campaigns aimed at studying flares
have been conducted and valuable information has been obtained
(e.g., Eckart et al. 2006a; 2008b; Yusef-Zadeh et al. 2006a, 2008,
2009; Marrone et al. 2008). It is found that each X-ray flare is
accompanied by an IR flare with zero time lag, but not vice versa.
The timescales of the variability are quite similar, typically
$\sim$ an hour (Eckart et al. 2006a; Yusef-Zadeh et al. 2006a). The
sub-millimeter flare is detected to lag the X-ray flare by
$110\pm17$ minutes (Yusef-Zadeh et al. 2008; Marrone et al. 2008;
Meyer et al. 2008). In the radio band, as shown by Figure 3, flares
at 43 and 22 GHz are detected and it is found that the peak of 22
GHz flare lags that of the 43 GHz by 20-40 minutes (Yusef-Zadeh et
al. 2006b, 2008).

Yusef-Zadeh et al. (2006b, 2008) argue that the time lag between 22
and 43 GHz, together with the rapid rise and slow decay of their
light curves and the polarization of the flare, all can be naturally
explained by a bubble of synchrotron-emitting electrons cooling via
adiabatic expansion. The initial rise of the flux is because of the
increase of the cross section of the blob when it is optically
thick. With the expansion, the blob becomes optically thin, and the
energy of electrons and the magnetic field both decrease, thus the
emitted flux decreases. Because the (self-absorption) optical depth
is a function of frequency and high-frequency emission becomes
optically thin first with the expansion, this results in the lag of
the 22 GHz peak compared to the 43 GHz one. This result tells us
that the flares are physically associated with ejection of blobs
from the accretion flow.

We would like to emphasize that the key point of this work is the
isolated blob. Such kind of mass ejection is different from the
continuous jet since it is episodic. In fact, so far we have not
detected a continuous jet in Sgr A*, as we point out earlier.

\subsection{An MHD model for the formation of episodic jets}

Interestingly, episodic mass ejection from accretion flow are quite
common and they are often called ``episodic jets'' (Fender \&
Belloni 2004). One evidence for episodic jets are the widely
detected knots in jets of active galactic nuclei (AGNs). These knots
are widely assumed to be due to collision of shells or blobs with
different velocities (so the AGNs jets we detect are mixture of
continuous and episodic mass outflows). In Galactic black holes,
episodic jets are most easily detected during the transition from
the hard X-ray spectral state to the soft X-ray spectral state. The
ejection of matter is often associated with brief and intense flares
in radio, IR, and X-ray bands, as in the case of Sgr A* (Mirabel \&
Rodriguez 1994; Mirabel et al. 1998; Fender et al. 1999; Hjellming
\& Rupen 1995; Corbel et al. 2002).

The characteristics of episodic ejections and steady, continuous
jets are distinguishable (see review by Fender \& Belloni 2004). The
feature of episodic jets is that their radio spectrum evolves
rapidly. It is initially optically thick and then becomes optically
thin. This is exactly what we have observed in the radio flares of
Sgr A*, but in contrast to continuous jets whose radio spectrum is
always optically thick. In addition, there is evidence that the
emission from ejected plasmoids is more highly polarized than the
emission from the continuous jets (Fender \& Belloni 2004). This is
again similar to the observations of Sgr A* in which the flare
emission is more polarized than in the quiescent state.

A natural question is then what causes the formation of episodic
jets. The formation of continuous jets has been intensively studied
and models have been proposed. It is believed to be associated with
the large-scale open magnetic field (e.g. Blandford \& Znajek 1977;
Blandford \& Payne 1982). However, the formation of episodic jets
remains an open question. Recently, by analogy with the coronal mass
ejection in the Sun, a magnetohydrodynamical model was proposed and
it is suggested to be associated with the closed magnetic fields
(Yuan et al. 2009). Figure 3 is the schematic picture of the model.
Loops of magnetic field emerge into the disc corona because of Park
instability. Since their foot points are anchored in the accretion
flow which is differentially rotating and turbulent, reconnections
and flares occur subsequently (e.g., Blandford 2002). The
reconnection process changes the magnetic field topology (see Fig.
3a). It also redistributes the helicity and stores most of it in a
flux rope floating in the disc corona, resembling what happens in
the Sun. Initially the flux rope is in equilibrium by the forces of
magnetic pressure and magnetic tension. The magnetic configuration
continues to evolve. The system eventually loses the equilibrium,
rapidly expelling the flux rope, and a long current sheet develops
behind the break-away flux rope, as shown by Fig. 3b. Magnetic
reconnection then occurs in the current sheet. The magnetic pressure
force becomes much stronger than the tension force, thus the flux
rope is propelled away from the accretion disc. Figure 4 shows our
calculation result of the velocity evolution of the ejected blob in
the case of Sgr A*. We can see that the blob is accelerated to
$0.8c$ within about half a hour. The readers are referred to Yuan et
al. (2009) for the details of calculation.

\begin{figure}[!h]
\includegraphics[scale=0.34,angle=0,trim=-20 -20 60 -10]{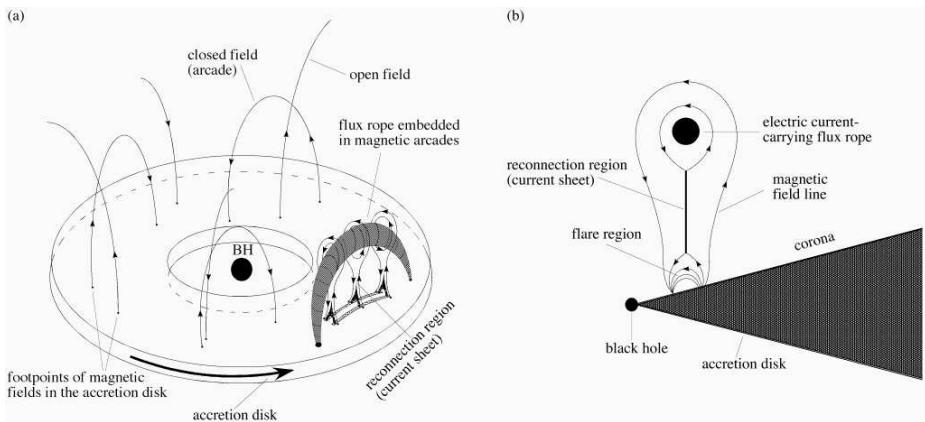}
\caption{Illustration of the formation of a flux rope and its
ejection. (a) Magnetic field emerges from the disc into the corona,
and a flux rope is formed, as a result of the motion of the
accretion flow and subsequent magnetic reconnection. (b) The flux
rope is then ejected, forming a current sheet. Magnetic reconnection
occurs subsequently, thus the magnetic tension becomes much weaker
than the magnetic compression. This results in the energetic
ejection of the blob. From Yuan et al. (2009).}
\end{figure}

\begin{figure}[!h]
\includegraphics[scale=0.5,angle=0,trim=-60 110 60 190]{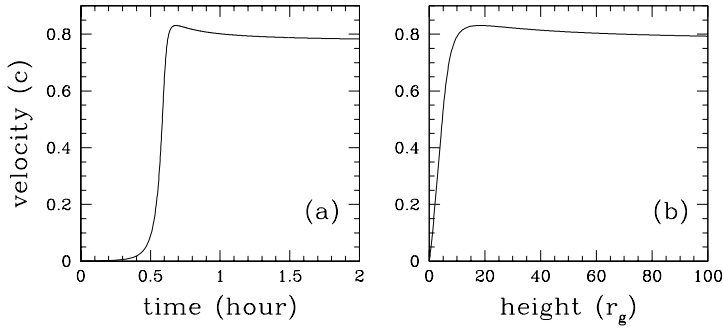}
\caption{The calculated velocity of the ejected blob from Sgr A* as
a function of time (a) and height (b). For comparison, the
light-crossing time across $r_g$ is $\sim 20$ s. From Yuan et al.
(2009).}
\end{figure}

\subsection{Toward uncovering the mystery of flares}

Several models have been proposed to explain the flares in Sgr A*.
They include shock acceleration in jet (Markoff et al. 2001),
magnetic reconnection in the accretion flow (Yuan, Quataert \&
Narayan 2004), turbulent acceleration in accretion flow (Liu et al.
2006a, 2006b), and ``hot spot'' model (Broderick \& Loeb 2005) in
which it is assumed that a localized region in the accretion flow
brighten during the flare due to some reason. Most of these works
are only phenomenological or they only focus on some aspects of the
flare.

As indicated by the observations of flares in Sgr A*, and our
knowledge of solar flare and coronal mass ejection in the Sun, the
ejection of blobs is often associated with flares. This implies that
the ejection and flare are basically two different appearance of the
same physical process. Therefore, we should be able to explain the
flares in Sgr A* based on the ``magnetic ejection'' scenario
introduced above. This work is still on-going (Yuan et al. in
preparation). An analogy with solar flare should be valuable but
significant differences are also expected due to the substantial
difference of physical conditions in solar corona and disk corona.
Important radiation should come from the relativistic electrons
accelerated in the current sheet during the reconnection process
(ref. Figure 3b). This radiation is likely to be responsible for the
IR and X-ray flares. When the ejected blob propagates in the
interstellar medium, the synchrotron emission from the blob itself
and/or from the shock formed in front of the blob should be
responsible for the radio and sub-millimeter flares. A time lag is
naturally expected between the IR (or X-ray) flares and the radio or
sub-millimeter ones. Because the magnetic field in the emission
region is ordered and the emission is optically thin, a high degree
of polarization is expected. As a comparison, for the quiescent
state, the emission is likely to come from accretion flow where the
magnetic field is more tangled and/or the emission is optically
thick, so the polarization degree is lower.

\section{Conclusion}

The supermassive black hole in our Galactic center, Sgr A*, is the
best laboratory to study the physics of black hole accretion. The
main observational constraints include the mass accretion rate and
the physical properties of the gas (temperature, density, and
specific angular momentum) at the Bondi radius, the spectral energy
distribution, and the polarization at radio wavebands. The
advection-dominated accretion flow satisfies all these constraints
and further is a well established theory for all low-luminosity
sources.

More attention has been paid to the flares of Sgr A* in recent
years. Many simultaneous multiwavelength campaigns have been
conducted. From these observations especially the time lag between
peaks of flares at different radio frequencies, we are now confident
that blobs are ejected from the accretion flow during the flare.
These episodic mass outflows are actually quite common in black hole
systems and usually called episodic jets. Their observational
appearances are different from those of continuous jets in several
important aspects such as spectrum and polarization. By analogy with
the coronal mass ejection in the Sun, an MHD model has been proposed
to explain the formation of episodic jets. Moreover, from the
observations of flares in Sgr A* and the Sun, we know that the
ejection of matter is often associated with them. In other words,
ejection and flares are intrinsically different appearances of the
same physical process. This implies that it is very promising that
we are able to understand the flares in Sgr A* along this line.

\acknowledgements This work was supported by the Natural Science
Foundation of China (grants 10773024, 10833002, 10821302 and
10825314), Bairen Program of Chinese Academy of Sciences, and the
National Basic Research Program of China (grants 2009CB824800 and
2006CB806303).


\end{document}